\documentclass{article}

\usepackage{microtype}
\usepackage{graphicx}
\usepackage{subfigure}
\usepackage{multirow}
\usepackage{booktabs} 
\usepackage{longtable}
\usepackage{hyperref}

\usepackage[accepted]{icml2025}

\usepackage{amsmath}
\usepackage{amssymb}
\usepackage{mathtools}
\usepackage{amsthm}

\usepackage[capitalize,noabbrev]{cleveref}

\theoremstyle{plain}

\theoremstyle{definition}

\theoremstyle{remark}

\usepackage[textsize=tiny]{todonotes}

\icmltitlerunning{Reporting standards for vision-language model studies}

\begin{document}

\twocolumn[
\icmltitle{Position: Restructuring of Categories and Implementation of Guidelines Essential for VLM Adoption in Healthcare}





\begin{icmlauthorlist}
\icmlauthor{Amara Tariq, Ph.D.}{yyy}
\icmlauthor{Rimita Lahiri, Ph.D.}{yyy}
\icmlauthor{Charles Kahn, M.D.}{comp}
\icmlauthor{Imon Banerjee, Ph.D.}{yyy}

\end{icmlauthorlist}

\icmlaffiliation{yyy}{Arizona Advanced AI hub, Mayo Clinic Arizona}
\icmlaffiliation{comp}{Department of Radiology, University of Pennsylvania}


\icmlkeywords{Machine Learning, ICML}

]



\printAffiliationsAndNotice{}  

\begin{abstract}
The intricate and multifaceted nature of vision language model (VLM) development, adaptation, and application necessitates the establishment of clear and standardized reporting protocols, particularly within the high-stakes context of healthcare. Defining these reporting standards is inherently challenging due to the diverse nature of studies involving VLMs, which vary significantly from the development of all new VLMs or finetuning for domain alignment to off-the-shelf use of VLM for targeted diagnosis and prediction tasks. In this position paper, we argue that traditional machine learning reporting standards and evaluation guidelines must be restructured to accommodate multiphase VLM studies; it also has to be organized for intuitive understanding of developers while maintaining rigorous standards for reproducibility. To facilitate community adoption, we propose a categorization framework for VLM studies and outline corresponding reporting standards that comprehensively address performance evaluation, data reporting protocols, and recommendations for manuscript composition. These guidelines are organized according to the proposed categorization scheme. Lastly, we present a checklist that consolidates reporting standards, offering a standardized tool to ensure consistency and quality in the publication of VLM-related research.
\end{abstract}

\vspace{-0.3in}
\section{Introduction}
\label{sc:intro}
Modern clinical practice is heavily dependent on the synthesis of information from multiple heterogeneous data sources, which is also crucial in medical image interpretation, where the integration of substantial clinical context is often necessary to ensure accurate and informed diagnostic decision-making. The ability to combine and interpret data from these diverse modalities allows holistic and precise assessments of patients' conditions. However, combining heterogeneous high-dimensional data while preserving the clinical knowledge necessary to support sensitive healthcare applications is an incredible challenge due to the data complexity, noise, semantic interoperability, bias, and ethical dilemmas~\cite{challenges_1, quinn2021trust}. 

Developers of deep learning systems have employed various fusion techniques (late, early, and joint fusion)~\cite{huang2020fusion} to join data and extract complementary information from multiple modalities. However, due to the supervised or semi-supervised nature of the learning, the models often learn spurious correlations and lack cross-modality semantic understanding of data~\cite{banerjee2023shortcuts}. Building on the success and widespread adoption of large language models (LLMs), recent research~\cite{radford2021learning,singh2022flava,li2021align,duan2022multi} has focused increasingly on extending self-supervised learning methodologies to multi-modal data which emphasizes the alignment and co-learning of diverse modalities through self-supervision, that not only helps in mitigating spurious correlations, but also leverages broader datasets by including unlabeled data. Furthermore, self-supervision~\cite{bommasani2021opportunities,gao2021simcse,chowdhery2023palm,radford2021learning,ramesh2021zero,singh2022flava} enables the broader application of models to multiple downstream tasks by avoiding the reliance on any single task for model training. This capability holds significant promise for healthcare where vast amounts of data are readily available, yet the primary challenge lies in the scarcity of labeled data for supervised learning. By leveraging self-supervised multi-modal learning, models can exploit unlabeled data, unlocking the potential for more scalable and effective solutions in clinical decision-making and healthcare analytics.

In recent literature, the Vision-Language Model (VLM) refers to a self-supervised co-learning paradigm that is designed to jointly process and understand visual (e.g., images or videos) and textual (e.g., text or speech) data, thus bridging the gap between modalities~\cite{radford2021learning,li2021align,singh2022flava,xu2021videoclip,you2023cxr,koleilat2024medclip}. Models trained under this paradigm are capable of reasoning across modalities and performing tasks that involve multi-modal semantic understanding. Even though design, development and application of VLM borrow heavily from LLM research, the learning objective, multi-modal data arrangement, model architectures and performance evaluation are more complex given the multi-modal nature of the problem. Given the research effort involving VLMs has exploded in the last few years~\cite{yildirim2024multimodal,van2024large}, particularly in finetuning and adapting of generic VLM for domain-specific healthcare tasks, there has been a growing need for standardized reporting and evaluation of these models. 

Our work is an effort towards updating traditional machine learning reporting standards~\cite{collins2024tripod+,tejani2024checklist} for VLMs to facilitate reproducibility in healthcare. \textbf{We argue that it is challenging to define reporting and evaluation standards for VLM studies without comprehensive categorization of various VLM research and development studies. Therefore, we present a comprehensive categorization scheme for VLM studies and then define reporting standards for healthcare applications.} Our reporting standards are organized as guidelines for sections of a standard research manuscript. They are also summarized as a checklist that may be adopted by publishers and peers while assessing VLM studies for publication to ensure standardized reporting style and reproducibility potential.

First, we outline the key reasons why reporting standards such as the TRIPOD+AI~\cite{collins2024tripod+} and the CLAIM~\cite{tejani2024checklist}, are unsuitable for VLMs. Afterwards, the present work: $(i)$ delineates clear categories of VLM development, finetuning, and applications to improve understanding, comparison, and reproducibility; $(ii)$ standardizes reporting of VLM design, multi-phase training, datasets, and performance evaluation; and $(iii)$ creates a simple checklist for peer evaluation of VLM studies. 

\vspace{-0.1in}
\section{Background}
\label{sc:back}
\emph{Multi-modal data fusion in healthcare:} Before the advent of VLMs, advances in multi-modal learning primarily relied on independently encoding textual and visual data using pretrained architectures, followed by fusing their respective embedding spaces for a specific downstream task in a supervised fashion~\cite{huang2020fusion,tariq2021patient}. In this approach, each modality (text and vision) was processed separately through domain-specific models, such as convolutional neural networks (CNNs) for images~\cite{li2023deep,zelaszczyk2023cross} and recurrent neural networks (RNNs) or transformers for text. The outputs from these models -- usually in the form of embeddings or feature vectors -- were then combined in a shared space, often through concatenation or projection, to enable the model to perform tasks like similarity-based retrieval, visual question answering, and data captioning. While simplistic methods like concatenation or projection, were effective for certain applications~\cite{zhao2019multimodal,jin2017multimodal}, they lacked the sophisticated cross-modal integration to model the complex interdependencies between the two modalities. Hence, such frameworks achieved some success for shared multi-modal embedding space generation~\cite{couairon2022embedding,wang2023mumic}, but lacked the capability of cross-modal reasoning within a unified framework required for complex tasks like visual prompting for data generation. In the healthcare domain, the combined embedding space can also be utilized for downstream classification and prediction tasks, particularly when the target task requires complementary information from both modalities. 

\emph{Vision-language model in healthcare:} Vision and textual data fusion based on concatenation or projection, lacked the nuanced alignment between the modalities that recent VLMs achieve. For latest VLMs, the multi-modal integration is not just a static fusion but an ongoing, dynamic process where vision and language are continuously aligned and integrated at multiple stages of the model. This enables the model to reason jointly over both modalities, fostering richer interactions and a context-aware understanding, which is critical for complex tasks such as visual question answering, image captioning, and cross-modal retrieval. In healthcare, the substantial impact of text supervision in vision models was markedly amplified with the release of CLIP (Contrastive Language-Image Pretraining) by OpenAI in early 2021 which employed contrastive learning framework to simultaneously train on both images and their corresponding textual descriptions, allowing for more flexible and scalable learning~\cite{you2023cxr,ghosh2024mammo,wang2022medclip,he2024pathclip}. Recently, CLIP has emerged as a foundational model and has catalyzed interest in various medical specialties, such as pathology and radiology. CLIP has also been adopted for disease diagnosis tasks in chest X-rays, organ segmentation and tumor detection in CT scans, breast density classification on mammograms. Text-conditional image generation using latent-diffusion model (LDM) has also been adopted in healthcare for high quality image generation where the power of generative diffusion models is used to create realistic, modality-specific images conditioned on textual descriptions~\cite{gu2022vector,xu2023versatile}. LDMs perform a diffusion process in a shared unified latent space that can learn to map both textual descriptions and visual features~\cite{gu2022vector,xu2023versatile}. The key advantage of LDMs is their ability to generate high-quality image data while operating in a compressed, unified latent space, significantly reducing computational complexity compared to traditional methods that worked directly in pixel space and allowing for the generation of images conditioned on text.

\emph{Limitations of traditional checklists:} Traditional checklists like TRIPOD+AI and CLAIM~\cite{collins2015transparent,mongan2020checklist, tejani2023updating} typically include essential components starting from model development and experimentation to  performance reporting that help ensure the rigor, reproducibility, and transparency of healthcare AI research. Given the extensive and detailed nature of checklists like TRIPOD+AI designed for traditional machine learning models, their direct adoption in reporting VLMs in healthcare can be challenging. This is primarily due to the inherent complexity of VLMs, particularly when dealing with multi-phase training processes (e.g., pretraining, domain alignment, and finetuning) and the diverse approaches for adopting pretrained VLMs (e.g., zero-shot, few-shot, and finetuning). The flexibility and variations in these processes add significant complexity to model development and evaluation pipelines that is not captured by standardized checklists designed for traditional end-to-end single-modal learning. For instance, while traditional checklists focus on standardized reporting for datasets, model architectures, and evaluation metrics, VLM studies often require more nuanced reporting of pretraining procedures, domain-specific cross-modal alignment, effect of different training objectives, analyzing bias at different training phases, report overlap with pretraining and task-specific finetuning datasets, etc. Due to this complexity, many recent healthcare VLM publications often report only selective components of the model architecture, training processes, and evaluation metrics. This selective reporting leads to challenges in replicating and adopting these models effectively, especially when trying to apply them across different healthcare settings or research studies, which are critical qualities for establishing trust in AI systems for healthcare. Thus, there is a growing need for standardized reporting frameworks tailored to the specific challenges of VLMs in healthcare.

\vspace{-0.1in}
\section{Categorization scheme for VLM studies}
\label{sc:categories}
\begin{figure}[htb!]
\vspace{-0.1in}
\begin{center}
\centerline{\includegraphics[width=\columnwidth]{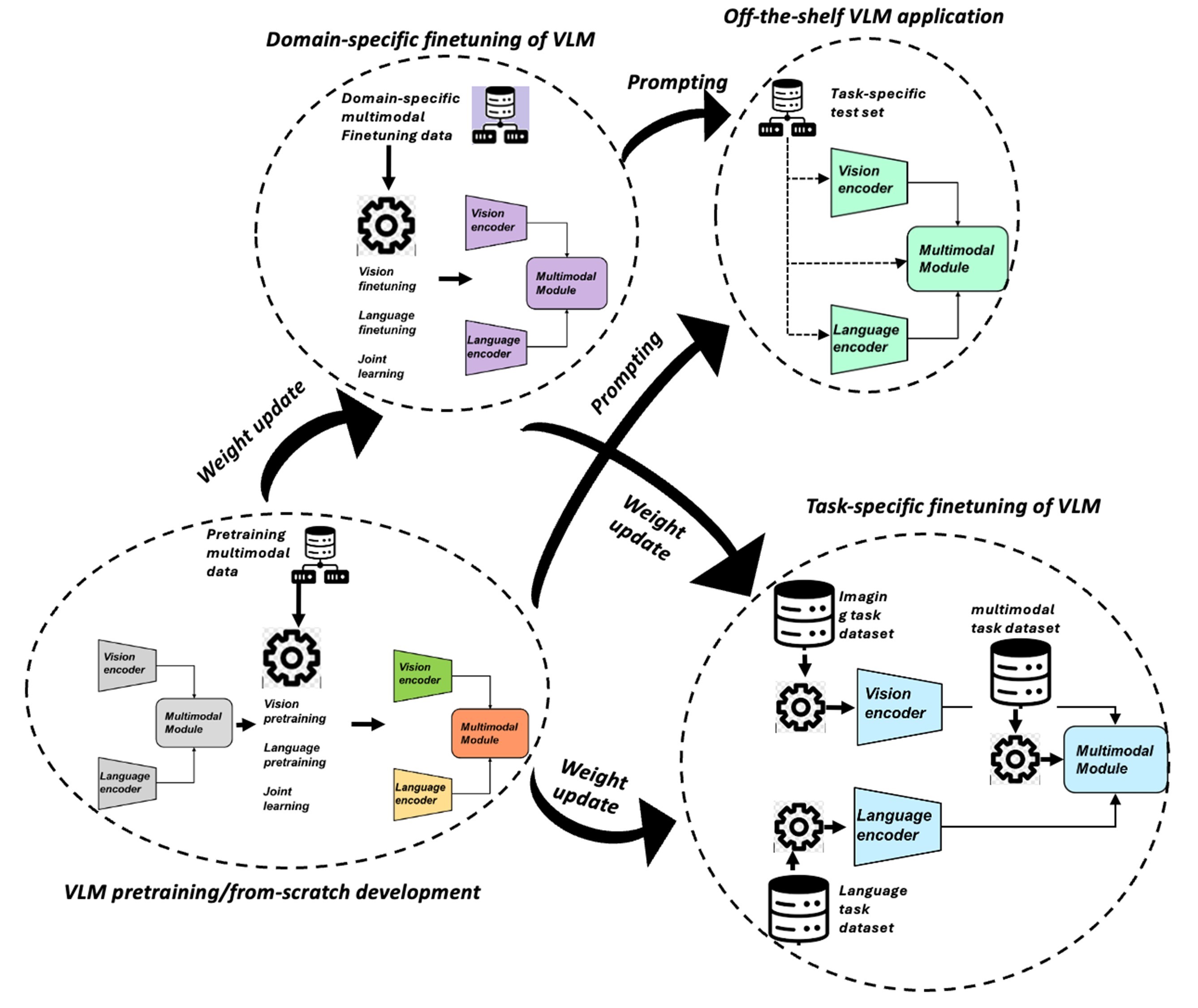}}
\caption{Conceptual framework of the proposed VLM study categorization where the arrows represent unidirectional inter-dependencies between the VLM categories.}
\label{fg:vlm_categories}
\end{center}
\vspace{-0.5in}
\end{figure}

We believe that comprehensive categorization of VLM research and development studies is a critical first step towards effective standardization of reporting guidelines for these studies and their reproducibility in healthcare. Based on the foundation of machine learning principles and insights drawn from the existing literature, we propose a categorization framework that divides VLMs into four distinct categories primarily derived from the key stages and strategies involved in the development, finetuning, and application. Figure \ref{fg:vlm_categories} shows this categorization and interdependencies between VLM development strategies.

\emph{(i) VLM pretraining:} In this category, we classify studies that propose a new VLM model with self-supervised pretraining where the objective is to model relationship between image and text modalities to learn a shared representation space. VLM pretraining allows encoding of both visual and textual information, and performs multiple downstream tasks that require cross-modal inputs. In literature, the objective functions associated with pretraining fall into four major categories: $(a)$ \emph{Masked image-region/token prediction:} Reconstruction of the original unmasked signal from its masked version for one or both modalities. MMBERT~\cite{khare2021mmbert} employed masked language modeling (MLM) for pretraining on medical visual question-answering (VQA) datasets by using image embedding as a pretext task. Xie et. al introduced MedIM~\cite{xie2023medim} with masked image modeling (MIM) as the pretraining strategy where masking was guided by radiology reports. FLAVA \cite{singh2022flava} was developed using a combination of masked modeling strategies for image, text, and multi-modal data; $(b)$ \emph{Contrastive learning:} Learning to maximize the similarity associated with paired vision-text tuples while minimizing the similarity between unpaired vision and text samples. CLIP framework proposed by Radford et. al \cite{radford2021learning}
is the seminal example of such learning. Recently, BioMedCLIP~\cite{zhang2023large} was developed by experimenting with different encoder architectures and model parameters for training a CLIP-like framework for the medical domain. $(c)$ \emph{Image-text matching:} Learning to predict cross-modality matches between images and text pairs~\cite{dou2022coarse,bao2022vlmo}. $(d)$ \emph{Hybrid strategies:} Combination of more than one pretraining objectives. Innovative combinations of masked modeling and image-text matching were used to pretrain MedVILL~\cite{moon2022multi} and M3AE~\cite{chen2022multi}. Prior works have also used combinations of masked prediction and contrastive alignment to leverage the complementary potential of these two pretraining strategies ~\cite{zhang2023multi,zhou2023advancing}.

\emph{(ii) Domain-specific finetuning:} While the VLM models pretrained on generic image-text pairs yield superior downstream task performance on natural images, the performance degrades drastically when these models are deployed on medical tasks due to the difference between natural and biomedical images (in terms of intensity, objects) and vast content variations between medical and web-scraped text.  We categorize studies that employ adaptation/finetuning of a generic VLM model weights for a specific domain with similar objective of aligning image and text modalities targeting more than one downstream task. For example, Huang et. al developed GLoRIA~\cite{huang2021gloria} by adapting ConVIRT\cite{zhang2022contrastive} through joint learning of attention-weighted image regions with words in the paired reports. Eslami et. al adapted PubMedCLIP\cite{eslami2023pubmedclip} from CLIP by finetuning on 80K medical image-text pairs from the ROCO dataset~\cite{pelka2018radiology}, but reported evaluations only on visual question-answering (VQA) tasks. 

\emph{(iii) Task-specific finetuning:} Given deep and complex domain knowledge required for the medical downstream tasks (e.g., categorization of tumor subtypes, prediction of a specific treatment outcome), VLM models trained or finetuned for the broader medical domain may fail to derive meaningful inference and need further task-specific finetuning. Such finetuning is usually performed on smaller data and updates the weights of the text and/or image encoder and the cross-modal alignment block to improve performance for a single or a set of targeted tasks. We categorize studies that aim to finetune pretrained VLMs (before or after the domain-specific finetuning) on a specific downstream task. For example, Mistretta et al.~\cite{mistretta2024re} employed incremental finetuning of BioViL~\cite{boecking2022making}; a publicly available state-of-the-art domain-specific VLM for chest X-ray; for multi-label disease diagnosis. We also consider studies that use partially frozen VLM encoders during finetuning within this category. For example, Vo et. al.~\cite{vo2024frozen} used a frozen CLIP model to initialize image and tabular data encoders and updated weights of a fusion classifier for breast cancer prediction that was fed concatenated representation from encoders. 

\emph{(iv) Prompting-based VLM studies:} In this category, the key focus is on leveraging pretrained or finetuned models to perform downstream tasks without modifying their architecture or weights. Instead, the model is prompted with text or images to perform tasks based on its existing knowledge, often through carefully designed or domain-specific prompts. For example, Mattejie et al.~\cite{mattjie2023zero} demonstrated zero-shot medical image segmentation capabilities of Segment Anything Model (SAM) trained on a combination of generic and medical datasets~\cite{kirillov2023segment}. Mishra et. al~\cite{mishra2023improving} demonstrated improvement in disease classification performance of pretrained CheXzero~\cite{tiu2022expert} through the use of positive and negative prompts.

\vspace{-0.1in}
\section{Reporting of model design and training}
\label{sc:design_training}
VLMs are built around distinct encoding modules for processing vision and language inputs separately, followed by co-learning modules that align the two modalities through various objectives which progressively shift from basic co-learning to more sophisticated forms of text guidance and eventually to full immersion, where the model integrates both modalities deeply during processing. Therefore, pretraining dataset and co-learning objectives are critical in determining the model's ability to generalize and perform well on downstream tasks. For example, alignment based objectives which focus on global matching may work better for image-text retrieval, report generation while generative modeling could be better suited for visual question answering (VQA), image retrieval; however, it is primarily an intuition and yet no theoretical derivation. 

The studies categorized within the pretraining or domain-specific finetuning must report the training objective as categorized in Section \ref{sc:categories} and report the comparative analysis with other objectives on a hold-out test set.  A VLM study may use multiple training objectives, for example, FLAVA~\cite{singh2022flava} simultaneously used masked language modeling and masked image modeling for pretraining language and image encoder respectively, and used masked multi-modal modeling to train multi-modal encoder. In such hybrid cases, a thorough ablation study is needed to understand the contribution of individual objectives. However, due to extensive computational and time requirements, often VLM pretraining is not systematically optimized and relies on empirical observations and practical experiment to determine the best configuration (e.g., which objectives should be prioritized, how much weight should be given to each objective, etc.). This empirical approach makes it difficult to understand why certain configurations work better than others, as it does not offer systematic insights into the individual contributions of each objective. Additionally, VLM pretraining studies should establish the merits for domain and task generalization in terms of more than one domain (included in and excluded from the pretraining dataset) and multiple downstream tasks similar and dissimilar to the pretraining objectives with prompting. Equally important is identifying the specific domains and tasks where the current pretrained model fails to generalize effectively and may need finetuning. Finally, the model weights should be made publicly available after pretraining or domain-specific finetuning to allow for verification of the VLM capabilities across a wide range of non-traditional use cases (more than one) from similar domains. However, due to legal restrictions surrounding patient privacy, models trained on private medical datasets are often not released, making such studies difficult to reproduce and hindering scientific progress and cause multiple VLMs to be pretrained on similar datasets with incremental improvement.

For domain-specific and task-specific finetuning, the first critical consideration is identifying a pretrained model from an overlapping domain (if available). It is essential to report the performance of the original pretrained model on the targeted domain and task using prompting, prior to any weight updates. Such reporting helps establish the necessity for expensive finetuning of the VLMs. If the author opts for a generic VLM instead of a publicly available domain-specific VLM, this choice must be justified with scientific reasoning. For example, studies still use the generic CLIP model for a targeted biomedical image-text generation task while MedClip, adopted for the biomedical domain~\cite{wang2022medclip}, is available.  Reporting of finetuning should include details of the objective function as well as information of the tuning dataset including patient-specific characteristics, e.g., age, gender, race, and optimized hyperparameters. After finetuning, the model should be evaluated on a held-out test set using relevant performance metrics. Additionally, it would be insightful to examine the semantic differences in the multi-modal joint embedding space before and after finetuning by multi-modal similarity assessment or exploring the cluster formed in the embedding space, as this could reveal how the model's understanding of the domain and task has evolved after adaptation to a domain or task~\cite{zheng2024fine,lai2310veclip}. 

For prompting-based VLM studies, the trend is to report the optimal performance for a specific task on a VLM without proper documentation of prompt optimization strategy or model versioning; however, it is essential to follow a clear and comprehensive set of guidelines to ensure transparency, reproducibility, and clarity for the prompting studies. In addition to specifying the pretrained VLM with correct versioning, it is also important to mention the source and the type of dataset the model was pretrained on (e.g., COCO~\cite{lin2014microsoft}, Conceptual Captions~\cite{sharma2018conceptual}, medical datasets~\cite{krishna2017visual}, etc.). Similar to finetuning, while adopting a domain-specific model, studies need to describe how the domain of the pretrained model overlaps with the target domain (e.g., medical, automotive, etc.) or if a generic model is used, studies need to justify this decision with scientific reasoning (e.g., lack of domain-specific models, better generalization capabilities, etc.). Details regarding prompting need to specify the generated context like zero-shot or few-shot, and explain if the prompt is template-based or dynamic, particularly for clinical reporting~\cite{kim2024chatgpt,rahsepar2023ai,javan2024gpt,hayden2024performance}. Given the current advancements in prompting optimization~\cite{nasiriany2024pivot,yu2025attention}, specifications of prompt generation are essential -- manual (human-crafted) or generated (e.g., based on a specific algorithm or task).

\vspace{-0.1in}
\section{Reporting of datasets}
\label{sc:datasets}
\begin{figure}[ht]
\vspace{-0.1in}
\begin{center}
\centerline{\includegraphics[width=\columnwidth]{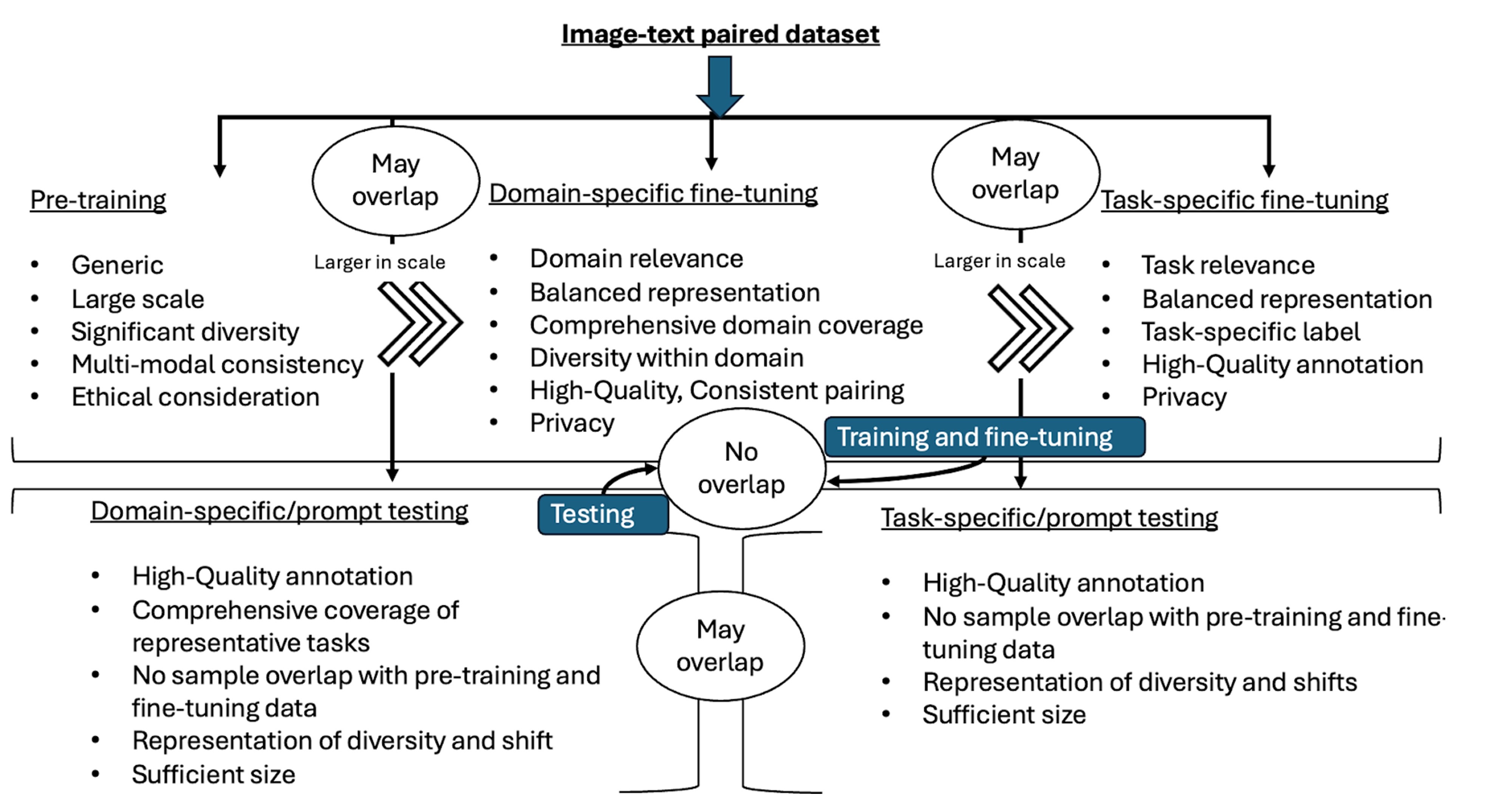}}
\caption{Conceptual categorization of VLM datasets based on our proposed study categorization.}
\label{fg:vlm_datasets}
\end{center}
\vspace{-0.3in}
\end{figure}

Based on our VLM study categorization, for unbiased validation, image-text paired dataset splits should be divided into $(i)$ pretraining set (used during VLM pretraining), $(ii)$ domain-specific finetuning set (used during domain-specific finetuning), $(iii)$ task-specific finetuning set (used during task-specific finetuning), $(iv)$ domain-specific testing set (used during testing of domain-specific VLM application), and  $(v)$ task-specific testing set (used during testing of VLM for specific task). VLM studies should report detailed characterization of each split. Figure \ref{fg:vlm_datasets} shows visual categorization of datasets and their primary selection criteria. Details are provided in the following section.   
\vspace{-0.1in}

\emph{Generic and large scale} -- The image-text paired dataset should be large enough to capture a wide variety of visual and textual patterns during VLM pretraining, enabling the model to generalize well across different domains. Common generic large-scale datasets for pretraining VLMs include COCO (2.5M)~\cite{lin2014microsoft}, Conceptual Captions (3.3M)~\cite{sharma2018conceptual}. Recently, large-scale medical datasets are also being used for pretraining VLMs - PMC-OA (1M)~\cite{lin2023pmc}, Visual Genome (100K)~\cite{krishna2017visual}. Given possible temporal similarity between studies of the same patient, medical datasets should be reported in terms of individual pairs and subjects.

\emph{Significant diversity} -- Dataset used for any VLM weight update, pretraining and domain- and task-specific finetuning, should be diverse. For medical dataset, diversity in visual representation (e.g., different sequence (T1, T2, DWI, ADC), different views (axial, sagittal, coronal), different contrast (venous, delayed)), population (patient age, gender, race, ethnicity, skin color, comorbidity) and downstream tasks (e.g., image interpretation, diagnosis, prediction, report generation, retrieval) should be considered. Temporal and geospatial diversities are also crucial for adoption of VLMs in healthcare. It is important to consistently report the diversity parameters to derive the generalization capabilities of pretrained or finetuned VLMs. 

\emph{Multi-modal consistency} -- Given the requirement for humongous scale, pretraining datasets are often curated automatically from web or existing data sources and thus include weak pairings of image-text relationships~\cite{desai2021redcaps,laurenccon2024obelics,garcia2023uncurated}. However, when dealing with finetuning with relatively smaller scale datasets for healthcare applications~\cite{birhane2021multimodal}, studies need to ensures that image-text pairings provides comprehensive information grounded in medical knowledge. For example, chest X-ray images should be paired with informative captions like ``Chest X-ray showing the right lung consolidation in the middle and lower lobes, consistent with pneumonia." instead of generic description ``Frontal view of female Chest X-ray”. Maintaining this consistency is crucial for adoption of VLM to perform complex medical tasks. Particularly for the testing datasets, manual verification for corruption and mislabeling is advised to ensure integrity of the dataset. The collection process for the medical data needs to be reported along with identification of experts, background and years of clinical experience.

\emph{Balanced representation, ethical considerations and privacy} -- are fundamental issues when working with multi-modal medical datasets, especially considering the sensitive nature of healthcare data and the strict regulations governing it's use. Ensuring compliance with privacy regulations like HIPAA in the U.S.~\footnote{https://www.hhs.gov/hipaa/for-professionals/privacy/laws-regulations/index.html} or GDPR in the European Union~\footnote{https://gdpr.eu/} is critical to safeguard patient confidentiality and uphold trust in the research and development process. Training and testing data should be carefully anonymized, including the pixel data (faces, pacemaker, device identification, unique anatomical features, etc.), to reduce re-identification risk for the patients. We must ensure that the dataset does not contain significant biases that could result in unfair predictions (e.g., gender or racial bias)~\cite{wan2024med}. It is important to maintain a transparent audit trail for healthcare data that tracks how data is collected, processed, and shared, and briefly report that in the study.

\emph{Domain and task relevance} --  Pretraining dataset can be generic but when considering domain and task-specific finetuning, the dataset must be directly representative to the domain and/or task. While task-specific datasets may focus on a single designated task~\cite{irvin2019chexpert,colak2021rsna}, domain-specific testing data should also be formatted in a way to conduct multiple task validation (e.g., segmentation, VQA, interpretation) to ensure that the model can be rigorously evaluated towards its generalization capabilities across the targeted domain. Relevance should be clarified in terms of clinical knowledge. 

\emph{Comprehensive coverage, diversity and shift} -- Similar to relevance, comprehensive domain or task coverage is needed for the finetuning and evaluation datasets. This can be achieved by considering data collection across modalities and institutions as well as diversity in patient population and care-team members. Strategic adjustment may be needed for collection of finetuning and testing datasets to capture data shifts, e.g., temporal changes across populations.  

\emph{Sufficient size} -- Though traditional sample size calculation is not applicable to non-linear VLMs, the sample sizes for finetuning and testing VLMs should take into account factors like model architecture, data diversity, and task complexity and can be determined through a combination of empirical testing, pilot experiments, and iterative refinement. However, current studies~\cite{ghosh2024mammo} are often limited by availability of public benchmark datasets which were often not particularly curated for VLM analysis. 

\emph{Data overlap} -- Ensuring that training and testing datasets are not overlapping is a crucial concern in VLM research, especially given the complexity of multiphase training and evaluation processes. Such data leakage can lead to overestimated performance and misrepresentation of  model's generalization ability. A model trained on a PubMed figure and caption dataset is then validated on a medical VQA task via prompting that includes test pairs from PubMed, is a classic example of this issue. Given that strict sample separation is infeasible due to the scale of the pretraining data, ensuring that the pretraining datasets do not overlap with the domain-specific testing datasets is a viable solution. Explicitly splitting data sets for task-specific finetuning is advisable such that no individual data point (image or text) is used in both the training and testing phases. 

Implementing strict dataset segregation, transparency in reporting, and external validation, will lead to accurate and fair reporting of true performance of VLMs in healthcare domain and accelerate their wide-spread adoption.
\vspace{-0.1in}
\section{Reporting performance}
\label{sc:performance}
\begin{figure}[htb!]
\vspace{-0.1in}
\begin{center}
\centerline{\includegraphics[width=1.1\columnwidth]{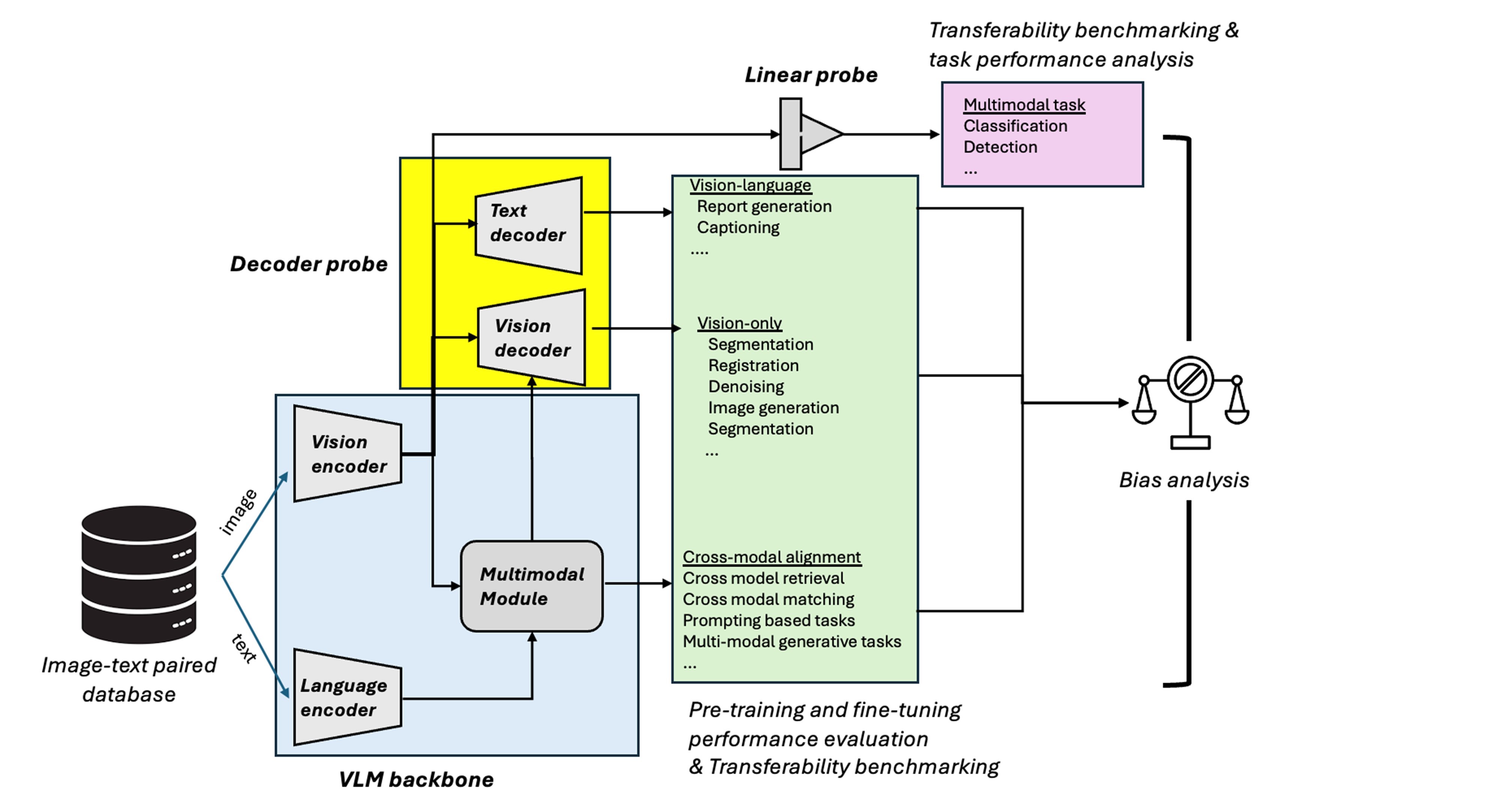}}
\caption{Conceptual diagram for VLM performance reporting.}
\label{fg:vlm_performance}
\end{center}
\vspace{-0.3in}
\end{figure}

Given the heterogeneous nature of VLM studies that involve multiple modalities, multiple encoding modules (vision encoder, text encoder, joint learning module) and multiphase training steps, standardization of performance evaluation is challenging. Traditional machine learning evaluation standards fail to accommodate VLM studies. We propose performance evaluation standards organized by the proposed categorization scheme of VLM studies (Figure \ref{fg:vlm_categories}). The guiding principle in defining these standards is that the claimed contributions of any VLM study must have corresponding performance evaluation. Table 1 in the appendix provides a list of standard metrics for reporting performance of the popular task categories mentioned in Figure \ref{fg:vlm_performance} (not exhaustive). Authors are encouraged to adopt alternative metrics if they can provide sound justification for their choices.

\emph{(i) VLM pretraining studies} should simultaneously benchmark the proposed pretraining strategy and transferability of the learned representation vectors on downstream tasks. 

\emph{Pretraining benchmarking:} Contrastive and matching-based pretraining should be evaluated with cross-modality matching tasks, e.g., zero-shot cross-modality retrieval, cross-modality reasoning. We posit that benchmarking must be balanced in terms of two modalities, for example, both image-to-text and text-to-image retrieval and matching performance should be reported.  For example,~\cite{wang2022medclip} proposed contrastive loss based alignment of vision and language representation in a joint embedding space, and evaluated on image-to-text and text-to-image retrieval. Generative modeling pretraining should be evaluated through cross-modality generative tasks like vision-to-language generation and language-to-vision generation (if allowed by the model architecture). For example,~\cite{singh2022flava} proposed masked language modeling (MLM), masked image modeling (MIM), and masked multi-modal modeling (MMM) as pretraining strategies and benchmarked with generative tasks like image-to-text generation.  

\emph{Transferability benchmarking:} As the major contribution of pretraining is to align the vision-language space to transfer well to a variety of downstream tasks, we posit that these studies must benchmark this transferability by experimenting with multiple downstream tasks. Benchmarking transferability and generalization are inherently challenging given widely varying characteristics of data and cross-modal relationships between data elements. We strongly recommend involving multiple data sources in downstream tasks to establish transferability of learned representation on a wide variety of scenarios. Downstream tasks may include VQA~\cite{guo2022images,eslami2023pubmedclip}, image/text classification, image/text generation~\cite{han2024integrating}, image-segmentation~\cite{roy2023sam}. 

\emph{(ii) Domain-specific finetuning studies} needs to benchmark $1)$ self-supervised finetuning and $2)$ transferability of learned representation in downstream tasks. Therefore, domain-specific finetuning needs two-pronged evaluation -- similar to pretraining studies -- and the same strategies and evaluation metrics can be applied but tailored to the targeted domain. For the domain of healthcare, it is imperative that the task is meaningful for clinical practice. For example, the general purpose image captioning task may be formulated as radiology image/volume interpretation and report generation~\cite{han2024integrating,gu2025radalign}. As pointed out in Section \ref{sc:datasets}, such benchmarking needs to ensure that there is no overlap between pretraining and domain-specific validation data. 

\emph{(iii) Task-specific finetuning and (iv) off-the-shelf applications}  evaluate performance of VLMs (or any of their components) on a specific downstream task with or without updating architecture or weights of the pretrained VLM. Therefore, machine learning performance reporting standards of (like TRIPOD+AI~\cite{collins2024tripod+,mongan2020checklist} and CLAIM~\cite{tejani2024checklist}) are applicable to these studies as they essentially use VLMs as deep learning encoders and decoders; only these VLMs are already pretrained to understand complex relationships between two data modalities thus facilitating performance on the downstream task. Given the multi-modality nature of VLMs, downstream tasks may include a vast variety broadly categorized into - $i)$ \emph{vision-only} --  where the encoders from the VLM are adopted for computer vision application and generates a image output, e.g., denoising, registration,  $ii)$ \emph{vision-language tasks} --  where text output is generated either only from image or image and text, e.g., image captioning; and $(iii)$ \emph{linear probing} -- where an MLP head is trained on top of a VLM backbone with the backbone's weights typically kept frozen for the multi-modal classification or prediction tasks. Given the popularity of LLMs, language-only tasks are usually achieved through LLMs. Compiling a comprehensive list of all downstream tasks may not be possible as VLMs are continuously being employed for novel applications.

\emph{Multi-phase bias analysis} -- VLMs rely on both image and language inputs, therefore in the pretraining phase, the bias can arise from the visual data (e.g., bias in image representation) and the language data (e.g., bias in language models). Bias can also emerge during cross-modality alignment. During the pretraining and finetuning benchmarking, thorough reporting of bias needs to be performed to measure whether the outcomes discriminate against a certain group (e.g., race, age, gender) based on statistical parity metrics, e.g., disparate impact, equalized odds. Disparate performance of the VLM at finetuning and pretraining stages should instigate further investigation of image and text encoders' bias. Bias also needs to be separately measured for each targeted task using - (i) group -- ensure that task performance is similar across different sub-groups based on demographics, geospatial and temporal characteristics. VLM that interprets chest CTs and generates recommendation for hospitalization for privately insured patients and outpatient visits for uninsured patients with similar diagnosis -- is a classic example of such bias. (ii) individual -- task output of similar instances, i.e., with semantically comparable visual and textual features, should be similar. For example, two chest X-rays showing the same severity of pneumonia must produce comparable interpretations when processed by the model, regardless of patient demographics.

\vspace{-0.1in}
\section{Discussion}
\label{sc:discussion}
The current digital healthcare paradigm is inherently multi-modal, where a variety of data types, such as images, vital signs, and clinical notes, are combined to assist in clinical decision-making. This fusion of heterogeneous data sources provides comprehensive insights into patient health, enabling healthcare providers to deliver more accurate diagnoses and personalized treatments. However, human experts face considerable challenges in interpreting and integrating such complex, multi-modal data streams for each individual patient, primarily due to scale and subtle differences in heterogeneous streams. While VLMs are gaining traction in research studies~\cite{ghosh2024mammo,blankemeier2024merlin} for their potential to assist in the continuous fusion of multi-modal data, their applications are often hypothetical. The clinical adoption of VLMs remains limited due to the lack of standardized frameworks and the absence of comprehensive benchmarking practices to assess their efficacy and reliability in real-world healthcare applications.
\vspace{-0.1in}

In this position paper, we proposed the new categorization scheme of the VLM studies for healthcare applications, developed reporting guidelines according to multi-phase development, and proposed a simplistic checklist for study review (Appendix). We established our position towards non-applicability of the traditional reporting standards for healthcare AI by highlighting limitations. Therefore, we proposed a new reporting guideline for reporting of datasets, methodology and performance of the VLMs based on the proposed categorization. Such guidelines will ensure standardization and the scientific rigor of VLM research, and enhance trustworthiness and clinical reliability, accelerating the adoption of VLMs in healthcare applications and improving patient outcomes in the process.

To facilitate better understanding, comparison, and reproducibility of VLM studies in healthcare, we propose a categorization framework that divides VLMs into four distinct categories based on the foundation of machine learning principles and insights drawn from the existing literature on VLMs -- VLM pretraining, domain-specific finetuning, task-specific finetuning, and prompting based studies. Distinct reporting guidelines for model design, training, datasets and performance evaluation are based on each VLM study category. We highlighted crucial considerations for each study type. For VLM studies, especially those focused on pretraining or domain-specific finetuning, it’s essential to report the training objectives and if multiple objectives are used (like FLAVA, which employs masked language and image modeling), an ablation study is necessary to understand the individual contributions of each objective. Public availability of model weights is essential to verify VLM capabilities across various use cases, though legal restrictions (e.g., patient privacy) often prevent the release of models trained on private medical data, hindering scientific progress. For prompting-based VLM studies, it’s critical to report the prompting context (zero-shot, few-shot, or template-based) and optimization (manual, automated) strategy and model versioning.

Based on proposed categorization, we also classified datasets used for VLM and highlighted the reporting criteria based on scale, diversity, representation and ethical considerations, domain and task relevance, domain coverage and data leakage. As best practices for VLM study reporting -- strict dataset segregation between different data classes should be maintained to ensure accurate testing; reporting of training and testing datasets should be transparent, and external validation sources should be used to ensure robust evaluation of VLMs. We recommend evaluation of each component of the VLM at different phases with the suggested metrics listed in Table 1 in Appendix and report performance for  -- $(a)$ pretraining; $(b)$ domain-Specific finetuning; $(c)$ task-Specific finetuning  and $(d)$ multi-phase bias. Based on the nature of VLM studies, all the performance may not be necessary to report but clear reasoning should be provided for comprehensiveness. These evaluation and bias analysis standards ensure fair and comprehensive performance measurement for VLMs in various healthcare applications, improving their reliability and trustworthiness in clinical settings. Finally, we summarized core points in checklist (Appendix Table 2) which can be adopted by authors or peer-review panels to ensure that the core components of the VLM reporting framework have been addressed. 

\emph{Limitations:} We developed the study categorization based on the current VLM literature published between 2020 - 2024 which may not be applicable to upcoming innovations or methodologies if they significantly change VLM training approaches. Being based on the machine learning foundation, the flexibility of the categorization framework allows for the possibility of extending the guidelines to accommodate these advancements as they arise. The list of performance metrics reported in Appendix is not exhaustive; it represents only the most commonly used metrics for each task. Authors are encouraged to adopt alternative metrics as long as they provide a sound justification for their choices.

\vspace{-0.1in}
\section{Alternate Position}
\label{sc:alternate}
While we strongly believe that establishing reporting standards for complex multi-modal VLM is important, we also propose an alternative approach, i.e., instead of creating a separate categorization scheme or developing a new checklist exclusively for VLM, we suggest extending existing machine learning reporting guidelines to account for the multi-stage nature of VLMs. This approach may leverage established frameworks while adapting them to the specific challenges of multi-modal learning, promoting consistency and broad applicability across the field.

\nocite{langley00}
\vspace{-0.2in}
\bibliography{vlm_checklist_references}
\bibliographystyle{icml2025}

\newpage
\appendix
\onecolumn
\section{Checklist}
We provide a list of recommended evaluation metrics in Table \ref{tab:evaluation}. In addition, we tabulated summary of our recommendations in the form of a checklist in Table \ref{tab:checklist}.

\begin{table}[htb!]
    \centering
    \begin{tabular}{|c|c|}\hline
       \textbf{Task} & \textbf{Evaluation Metrics}\\\hline
       Cross-modal retrieval  & Recall@K, Precision@K (K based on dataset size and task complexity)\\\hline
        Classification/Detection & Sensitivity and specificity, AUROC\\
         & Precision and Recall, F-score; alongwith confidence intervals\\\hline
        Language generation & Quantitative score in language space: METEOR, BLEU, etc.\\
         & Quantitative score in embedding space: BERT score, G-eval, etc.\\\hline
        Image generation & Automated: SNR, PSNR, FiD~\cite{heusel2017gans}, Inception Score~\cite{salimans2016improved}\\
         & Manual: Verification of synthetic images by domain experts\\\hline
        Segmentation & Dice coefficient, Jaccard coefficient (`IoU')\\\hline
        Prediction & Concordance index (C-index), time-dependent ROC \\\hline
        Bias analysis & Pretraining and finetuning: Statistical parity, disparate impact, equalized odds\\
         & Task evaluation: Group and individual fairness\\\hline
    \end{tabular}
    \caption{List of recommended evaluation metrics}
    \label{tab:evaluation}
\end{table}


\begin{longtable}{|l|l|l|}
\hline
\textbf{Section}              & \textbf{Subsection}                                                                                       & \textbf{Description}   \\ \hline
\endfirsthead
\endhead
Title                         &                                                                                                           & \begin{tabular}[c]{@{}l@{}}Core concept: Mention imaging and language data elements included, the \\ clinical domain and scope of the VLM and/or its application.\end{tabular}                                                                                                                                                                                                                                                                                                                                                                                                                                                                                                                                                                                         \\ \hline
\multirow{5}{*}{Abstract}     & Objective                                                                                                 & \begin{tabular}[c]{@{}l@{}}Brief description of broad aim of the study (e.g., development of improved \\ VLM, alignment of pretrained VLM with a specific domain or task, off-the-\\ shelf prompting) and its anticipated impact.\end{tabular}                                                                                                                                                                                                                                                                                                                                                                                                                                                                                                                         \\ \cline{2-3} 
                              & \multirow{2}{*}{\begin{tabular}[c]{@{}l@{}}Materials and \\ methods\end{tabular}}                         & \begin{tabular}[c]{@{}l@{}}Study Design: Identify all involved categories of study design (details in \\ Methods section)\end{tabular}                                                                                                                                                                                                                                                                                                                                                                                                                                                                                                                                                                                                                                 \\ \cline{3-3} 
                              &                                                                                                           & \begin{tabular}[c]{@{}l@{}}Data description: Brief statistical description of pretraining/finetuning/task-\\ specificbdata and characteristics.\end{tabular}                                                                                                                                                                                                                                                                                                                                                                                                                                                                                                                                                                                                          \\ \cline{2-3} 
                              & Conclusion                                                                                                & \begin{tabular}[c]{@{}l@{}}Description of merits of the study in terms of performance and achievement of\\ stated objective\end{tabular}                                                                                                                                                                                                                                                                                                                                                                                                                                                                                                                                                                                                                               \\ \cline{2-3} 
                              & Public access                                                                                             & \begin{tabular}[c]{@{}l@{}}Public access policy and (if applicable) links to data, code repository, model \\ weights, web-based demonstrations, etc.\end{tabular}                                                                                                                                                                                                                                                                                                                                                                                                                                                                                                                                                                                                      \\ \hline
\multirow{3}{*}{Introduction} & Background                                                                                                & \begin{tabular}[c]{@{}l@{}}Description of clinical challenge: intended clinical use-cases of VLM, clinical \\ impact\end{tabular}                                                                                                                                                                                                                                                                                                                                                                                                                                                                                                                                                                                                                                      \\ \cline{2-3} 
                              & Past research                                                                                             & Current state-of-the-art for medical domain/task with and without VLM                                                                                                                                                                                                                                                                                                                                                                                                                                                                                                                                                                                                                                                                                                  \\ \cline{2-3} 
                              & \begin{tabular}[c]{@{}l@{}}Proposed \\ approach\end{tabular}                                              & \begin{tabular}[c]{@{}l@{}}Study Contributions: Hypothesis, core contributions, selection of data elements, \\ downstream clinical task(s)\end{tabular}                                                                                                                                                                                                                                                                                                                                                                                                                                                                                                                                                                                                                \\ \hline
                              & Design                                                                                                    & \begin{tabular}[c]{@{}l@{}}Study design and methodology: Report for all phases of the study (listed below)\\ 1- VLM pretraining\\ 2- Domain-specific finetuning\\ 3- Task-specific finetuning\\ 4- Prompting-based/off-the-shelf use of VLM\end{tabular}                                                                                                                                                                                                                                                                                                                                                                                                                                                                                                               \\ \hline
                              & Data                                                                                                      & \begin{tabular}[c]{@{}l@{}}Report data statistics for all study phases separately\\ Study population: Inclusion/exclusion criteria, dataset size, patient and \\ acquisition (if applicable) characteristics highlighting diversity\\ Multi-modal data: Relationship between data elements (paired/unpaired), \\ groundtruth annotation and verification (if applicable)\\ Data splits: Data subset sizes, level of splitting (patient-, institution-,  study-, \\ source- level split), overlap between splits\\ Preprocessing details: Vision/language preprocessing techniques \\ (normalization, augmentation, truncation/padding, etc.)\\ Deidentification details: if applicable\\ Missing data: Strategies to handle missing data (vision/language)\end{tabular} \\ \hline
\multirow{4}{*}{Methods}      & Model                                                                                                     & \begin{tabular}[c]{@{}l@{}}Model architecture details: vision/language encoder/decoder, fusion/joint \\ learning strategy, probing methodology\end{tabular}                                                                                                                                                                                                                                                                                                                                                                                                                                                                                                                                                                                                            \\ \cline{2-3} 
                              & \begin{tabular}[c]{@{}l@{}}Pretraining or \\ self-supervised \\ finetuning\\ (if applicable)\end{tabular} & \begin{tabular}[c]{@{}l@{}}Learning task(s): Pretraining objectives for vision/language encoders, \\ cross-modality alignment/joint learning objective\\   - Full/partial model weights update\\   - Loss formulation\\   - Computational resources and time\\   - Software packages and libraries (include version)\end{tabular}                                                                                                                                                                                                                                                                                                                                                                                                                                      \\ \cline{2-3} 
                              & Downstream task                                                                                           & \begin{tabular}[c]{@{}l@{}}Downstream task details:\\  - Vision only: Detection, classification, segmentation, registration, \\ clustering, retrieval, etc.\\  - Language only: Generative, extractive, classification, clustering, \\ retrieval, etc.\\  - Vision-language: Cross-modal retrieval, cross-modal generation, etc.\end{tabular}                                                                                                                                                                                                                                                                                                                                                                                                                          \\ \cline{2-3} 
                              & \begin{tabular}[c]{@{}l@{}}Prompting details\\ (if applicable)\end{tabular}                               & \begin{tabular}[c]{@{}l@{}}Provide prompt text including examples used for few-shot training\\ or in-context learning\\ Highlight prompt optimization strategies\end{tabular}                                                                                                                                                                                                                                                                                                                                                                                                                                                                                                                                                                                          \\ \hline
\multirow{2}{*}{Evaluation}   & \begin{tabular}[c]{@{}l@{}}Pretraining or\\ self-supervised\\ finetuning\\ (if applicable)\end{tabular}   & \begin{tabular}[c]{@{}l@{}}Performance on pretraining/finetuning objective task evaluated on \\ held-out dataset, e.g., cross-modality retrieval or generation\\ Recommendation: Evaluation should be performed for both vision and\\ language tasks in a balanced fashion, e.g., reporting of image-to-text and\\ text-to-image retrieval performance\end{tabular}                                                                                                                                                                                                                                                                                                                                                                                                    \\ \cline{2-3} 
                              & \begin{tabular}[c]{@{}l@{}}Downstream \\ task(s)\end{tabular}                                             & \begin{tabular}[c]{@{}l@{}}Performance reporting with recommended metrics (Table XX) with \\ uncertainty measures like confidence interval;\\ Report bias analysis results for minority subgroups (particularly \\ recommended for the studies in the domain of healthcare)\end{tabular}                                                                                                                                                                                                                                                                                                                                                                                                                                                                               \\ \hline
Discussion                    &                                                                                                           & \begin{tabular}[c]{@{}l@{}}Provide scientific observations and highlight limitations;\\ Clinical impact in light of performance (if applicable),\\ Comparative performance for vision and language tasks,\\ Biases and privacy concerns, Uncertainty in prompt-based tasks\end{tabular}                                                                                                                                                                                                                                                                                                                                                                                                                                                                                \\ \hline
\caption{Checklist for standardization of peer-review process of VLM studies}
\label{tab:checklist}
\end{longtable}


\end{document}